\documentstyle[prd,aps]{revtex}
\begin{document}
\input epsf.tex

\title{A Solution to the Hierarchy Problem with an Infinitely Large Extra
Dimension and Moduli Stabilisation}
\author{Z. Chacko and Ann E. Nelson}
\address{ Department of Physics, University of Washington, Seattle,
Washington, WA 98195, USA.}
\maketitle
\preprint{UW/PT-99/28}

\begin{abstract}

We construct a class of solutions to the Einstein's equations for
dimensions greater than or equal to six. These solutions are characterized
by a non-trivial warp factor and possess a non-compact extra dimension. We
study in detail a simple model in six dimensions containing two four
branes. One of each brane's four spatial directions is compactified. The
hierarchy problem is resolved by the enormous difference between the warp
factors at the positions of the two branes, with the standard model fields
living on the brane with small warp factor.  Both branes can have positive
tensions. Their positions, and the size of the compact dimension are
determined in terms of the fundamental parameters of the theory by a
combination of two independent and comparable effects---an anisotropic
contribution to the stress tensor of each brane from quantum fields living
on it and a contribution to the stress tensor from a bulk scalar field.
One overall fine tuning of the parameters of the theory is required
---that for the cosmological constant. 
 
\end{abstract} \pacs{}

\vskip2pc

\section{Introduction}

The work of Arkani-Hamed, Dimopoulos and Dvali (ADD)  \cite{add} and of
Randall and Sundrum (RS) \cite{rs} has stimulated interest in explaining
the observed weakness of gravity (the ``hierarchy problem'') using extra
dimensions. The ADD solution requires the new dimensions to have
finite  but 
 large volume, 
which introduces a new hierarchy between the volume of the
compact dimension and the fundamental scale of the theory\footnote{For
 examples of  theories which naturally have finite but exponentially
large volume for the additional dimensions see ref.~\cite{ck}. For earlier work on large extra dimensions and/or a low
quantum gravity scale see refs.~\cite{early}.}. RS proposed
instead five dimensional spacetime with curvature comparable to the
fundamental scale, and showed that a massless graviton can be localized to
a 3+1 dimensional hypersurface known as the ``Planck Brane''.  In their
setup the extra dimension may be taken infinitely large, and four
dimensional general relativity still agrees to high precision with long
distance experimental measurement. The weakness of observed gravity is
explained provided the standard model fields are localized to a 3+1
dimensional ``TeV Brane'', where the graviton wave function is small.  Due
to the exponential fall-off of the graviton wave function away from the
Planck brane, the distance between the TeV brane and the Planck brane does
not need to be large in units of the fundamental scale.  Randall and
Lykken  (RL) \cite{rl} have shown that an infinite fifth dimension is
experimentally quite consistent with such a resolution of the hierarchy
problem. 

In such a picture, it is necessary to introduce dynamics which determines
the location of the TeV brane relative to the Planck brane. If this
interbrane distance is not fixed, it becomes a massless modulus which
leads to unnacceptable cosmology \cite{cosmo} and experimental
consequences. Goldberger and Wise showed that adding a scalar field which
propagates in the bulk and has a source on the branes is sufficient to fix
this distance \cite{gw}. Several other suggestions have been made for
bulk dynamics to fix the extra dimensional configuration \cite{stabilize}. 

In this paper we first construct a class of solutions to the Einstein's
equations for dimensions greater than or equal to six. These solutions are
characterized by a non-trivial warp factor and possess a non-compact extra
dimension.  We then study in detail a simple model in six dimensions
containing two four branes that employs a metric of this form to address
the hierarchy problem. One of each brane's four spatial directions is 
compactified on a circle of small radius. The hierarchy problem is
resolved as in the RS and RL models by the enormous difference in the 
warp factors at the positions of the two branes, with the standard model 
fields living on the brane with small warp factor. 

In this model the positions of the branes, and hence the magnitude of the
hierarchy, are determined by the combination of two independent effects.
The first is an anisotropic contribution to the stress tensor of each
brane arising from the quantum effects of fields localised to it. The
second is a contribution to the stress tensor from a scalar bulk field as
in the model of Goldberger and Wise. These effects are naturally
comparable in size and together can  yield a sufficiently large value of
the brane spacing to solve the hierarchy problem without fine tuning of
parameters.

The reason for the first effect is that the theory contains a compact
dimension, in addition to the noncompact dimension $r$.  The size of the
compact dimension is in general an $r$ dependent function, which is
determined from the Einstein's equations. We argue that in general the
component of the brane tension in the compact dimension will depend on its
size, due to the quantum effects of fields localized to the brane.  Then a
consistent solution to Einstein's equations will fix each brane location
at a particular value of $r$. However in this model this effect by itself
does not give rise to a large hierarchy without fine tuning. Nevertheless
when combined with the effect of a bulk scalar field on the geometry 
it is possible to realise a large hierarchy without fine tuning of 
parameters.  

Both branes in the theory can have positive tensions and the solution is
free of singularities where general relativity might break down. One
overall fine tuning of the parameters in the theory is required to adjust
the four dimensional cosmological constant to zero.

\section{The Ansatz, Equations of Motion and Solutions}

We begin by looking for solutions to the Einstein equations in D
dimensions, where D is greater than or equal to six, in the presence of a
constant background bulk cosmological constant. We assume all sources other
than the bulk cosmological constant are restricted to subspaces of lower
dimension. Hence our
approach will be to first solve the equations of motion in the bulk to
obtain solutions with a number of constants of integration that can then
be adjusted to find solutions for various
 boundary conditions.

The action in the bulk is 

\begin{equation}
S = \int d^D x \sqrt{-G} (2 M_*^{D-2} R - \Lambda_B)\ .
\end{equation}
  
We label a general coordinate by $x^M$ where $M$ runs from 0 to (D-1).  We
restrict our search to metrics of the simple form

\begin{equation}
ds^2 = f(z) \eta_{\mu \nu} dx^{\mu} dx^{\nu} + s(z)dy^2 + dz^2
\end{equation} 
where $\mu$ and $\nu$ run from 0 to D-3. The remaining two coordinates are
labelled by $y$ and $z$.  Here the warp factors $f$ and $s$ are assumed to be
functions only of $z$. 
 
The Einstein's equations in the bulk take the form

\begin{equation}
2 M_*^{D-2} (R_{MN} - g_{MN}R) = -\frac{1}{2} g_{MN} \Lambda_B\ .
\end{equation} 

The nontrivial components of this equation are 

\begin{equation}
\frac{1}{2} f'' (3-D) -\frac{1}{2} f \frac{s''}{s} 
+ f' \frac{s'}{s} \frac{(3-D)}{4} + \frac{1}{4}f \left(\frac{s'}{s}\right)^2
- f\left(\frac{f'}{f}\right)^2 \left(\frac{D^2-9D+18}{8}\right) 
= -\alpha^2 f \frac{(D-2)^2+(D-2)}{8} 
\end{equation}

\begin{equation}
\label{eq:feqn}
\frac{1}{2}\left[(D-2)\frac{f''}{f} + 
\frac{(D-2)(D-5)}{4}\left(\frac{f'}{f}\right)^2\right]
= \alpha^2 \frac{(D-2)^2+(D-2)}{8}
\end{equation}

\begin{equation}
\label{eq:seqn}
\frac{(D-2)(D-3)}{8}\left(\frac{f'}{f}\right)^2 +
\frac{D-2}{4}\frac{f'}{f}\frac{s'}{s} = 
\alpha^2 \frac{(D-2)^2+(D-2)}{8}
\end{equation}
where $\alpha$ is defined by

\begin{equation}
\alpha^2 \frac{(D-2)^2+(D-2)}{8} = -\frac{\Lambda_B}{4M_*^{(D-2)}}\ .
\end{equation}

To solve these equations note that we can rewrite eqn.~(\ref{eq:feqn}) in the form

\begin{equation}
\frac{1}{2}\left[(D-2)\left(\left(\frac{f'}{f}\right)'+\left(\frac{f'}{f}\right)^2\right)
 + \frac{(D-2)(D-5)}{4}\left(\frac{f'}{f}\right)^2 \right]  
= \alpha^2 \frac{(D-2)^2+(D-2)}{8}\ .
\end{equation}

This has the form of a first order differential equation for 
$\frac{f'}{f}$. This differential equation is straightforward to 
solve, and we can then obtain $f$ itself by performing a simple 
integration. We then use the result obtained for $f$ in
eqn.~(\ref{eq:seqn}) and
the problem of determining $s$ also then reduces to performing a
simple integral. The results  are 

\begin{equation}
\label{eq:f}
f = f_0e^{\alpha
z}\left[1-ce^{-\frac{D-1}{2}{\alpha}z}\right]^{\frac{4}{D-1}}
\end{equation}

\begin{equation}
\label{eq:s}
s = s_0 e^{\alpha z}\left[1-ce^{-\frac{D-1}{2}{\alpha}z}\right]^{-2\frac{(D-3)}{D-1}}
\left[1+ce^{-\frac{D-1}{2}{\alpha}z}\right]^2\ .
\end{equation}
  
Here $f_0$, $s_0$ and $c$ are constants to be determined by boundary
conditions. In the limits of vanishing $c$ and infinite $c$ we recover the
usual anti deSitter (AdS) metric. It is easy to see that in fact for any values
of these constants the warp factors $f$ and $s$ change very rapidly as
function of $z$. In particular there are always values of $z$ where they
are changing exponentially quickly. This suggests that these metrics are
good candidates for a solution to the hierarchy problem. 

It is possible to use the class of metrics above to find solutions to the
Einstein equations for various source configurations and
geometries\footnote{Note that these solutions are  coordinate
transformations of the bulk solutions found in ref.~\cite{cp}. }. In
the next section we exhibit a potential solution to the hierarchy problem
based on these metrics.

\section{The Model}

\subsection{The Metric}

In this section we limit our interest to a solution where the fifth
dimension $y$ is compact and corresponds to an angle in the higher
dimensional space. We relabel $y$ by $\phi$ in this section and hereafter to
emphasize its angular character. The angle $\phi$ runs from zero to
$2\pi$. We allow the coordinate $z$ to be non-compact and run from
zero to infinity. it corresponds to a `radius' in the higher dimensional
space. We relabel it by $r$ to emphasize its radial character.

The geometry of our model consists of two four branes localized in the
higher dimensional space at different values of $r$. Their positions are
specified by the equations $r=a$ and $r=b$ where $a<b$. They can
therefore
be thought of as being similar to the surfaces of two infinitely long
`concentric cylinders' in the higher dimensional space, with the regular
four dimensions parallel to the common axis of the cylinders, the fifth
dimension going around the surface but perpendicular to the axis, and the
sixth dimension being the radius.  (This intuitive picture does
not
account for the fact that the space is curved.)  The standard model fields
are localized to the brane at $r=a$.  The hierarchy problem will be
resolved by the enormous difference between the values of the warp factor
at $r=a$ and $r=b$. 

Because the coordinate $\phi$ is compact the four brane appears as a
three brane at sufficiently long length scales.
 
The branes divide the space into three distinct sections; $0< r < a$
,  $ a< r < b $, and $r > b$.  In general there is no reason for the bulk
cosmological constants in these three sections to be the same since
the branes
may be separating different phases of the theory. In what follows we will
assume that they are different and will associate the three regions with
three different values of $\alpha$; $\alpha_1$, $\alpha_2$ and $\alpha_3$
respectively. 

The solutions of the Einstein equations in the three regions will be
of the form of equations (\ref{eq:f}) and (\ref{eq:s}) 
but with different values of the 
constants $f_0$, $s_0$ and $c$. We will give these constants an additional
subscript $i$, where $i$ runs from 1 to 3, in order to differentiate them in
the three different regions. 

Now the constant $c_1$ is fixed to be -1 by the requirement that the
solution be non-singular at the origin. This requirement also fixes the
value of $s_{01}$ to be $(2^{16/5}/(25\alpha^2))$. 
To see that this choice does indeed smooth out
the singularity at the origin we first examine the behavior of the 
the functions $f(r)$ and $s(r)$  as $r$ tends to 0,

\begin{eqnarray}
f(r) &=& const\;+ O(r^2) \\
s(r) &=& r^2 + O(r^4)\ .
\end{eqnarray} 

We then go to the `cartesian' coordinate system which is smooth at the
origin. 

\begin{eqnarray}
x' &=& r\;\;  \cos\;\phi \\
y' &=& r\;\;  \sin\;\phi\ .
\end{eqnarray}

It is straightforward to verify that in this coordinate system the
components of the metric and their first and second derivatives (which go
into the Ricci tensor) are smooth at the origin, showing that there is
no singularity there.
 
The requirement that the metric be bounded at infinity determines that the
space outside $r=b$ be AdS. This corresponds to setting $c_3$
to infinity while keeping the products $c_3 f_{03}$ and $c_3 s_{03}$
finite. We relabel these products by $f_3$ and $s_3$ respectively.

We are now in a position to write down the forms of the solutions
in the three regions. For $ r< a$,

\begin{equation}
f = f_{01}e^{\alpha_1 r }[1 + e^{-\frac{5}{2}{\alpha_1}r}]^{\frac{4}{5}}
\end{equation}

\begin{equation}
s = s_{01} e^{\alpha_1 r}[1 + e^{-\frac{5}{2}{\alpha_1}r}]^{\frac{-6}{5}}
[1 - e^{-\frac{5}{2}{\alpha_1}r}]^2\ .
\end{equation}

For $b > r > a$,

\begin{equation}
f = f_{02}e^{\alpha_2 r }[1 - c_2 e^{-\frac{5}{2}{\alpha_2}r}]^{\frac{4}{5}}
\end{equation}

\begin{equation}
s = s_{02} e^{\alpha_2 r}[1 - c_2 e^{-\frac{5}{2}{\alpha_2}r}]^{\frac{-6}{5}}
[1 + c_2 e^{-\frac{5}{2}{\alpha_2}r}]^2
\end{equation}

For $r > b$,

\begin{equation}
f = f_3e^{-\alpha_3 r}\ ,
\end{equation}

\begin{equation}
s = s_3 e^{-\alpha_3 r}\ .
\end{equation}

The constant $f_{01}$ is determined by normalizing the warp factor $f$ to 1
at the position of the visible brane. The constant $s_{01}$ was determined
earlier. All the other constants above as well as the positions of
the branes $a$ and $b$ must be determined by matching across the branes.

We now write down the action for the branes. 

\begin{equation}
S_b = \int dx^M \sqrt{-\bar G}(\delta(r-a)[L_{M1} - \bar{\Lambda}_1] +
\delta(r-b))[L_{M2} - \bar{\Lambda}_2])\ .
\end{equation}

Here $\bar G$ is the determinant of the metric tensor in the five 
dimensional subspace, $\bar{\Lambda}_1$ and $\bar{\Lambda}_2$
are the cosmological constants on the two branes, and $L_{M1}$ and
$L_{M2}$ are the Lagrangians of matter fields localized to the branes.
  
The stress tensor for each brane has the form 

\begin{equation}
T_{AB} = {T^{\Lambda}}_{AB} +  
<{T^M}_{AB}>
\end{equation}
where ${T^{\Lambda}}$ is the contribution from the cosmological constant
and $<T^M>$ the expectation value of the stress tensor of the matter
fields living on the brane. We will be working in the semi-classical 
limit, treating gravity completely classically but accounting for the
quantum effects of matter localized to the branes. 

Assuming the matter on the brane is in its ground state,  
$T_{AB}$ is constrained by four dimensional Lorentz invariance
to be of the form 

\begin{eqnarray}
\label{eq:stress}
\frac{8 \pi}{2M_*^4}T_{AB} = -\left(\begin{array}{cccccc}
-\beta^2f & 0 & 0 & 0 & 0 \\
 0 & \beta^2f & 0 & 0 & 0 \\
 0 & 0 & \beta^2f & 0 & 0 \\
 0 & 0 & 0 & \beta^2f & 0 \\
 0 & 0 & 0 & 0 & \gamma^2s \end{array} \right)\ ,
\end{eqnarray}
where $\beta^2$ and $\gamma^2$ are constants associated with each brane
that we require to be positive. The form of ${T^{\Lambda}}$ is more
constrained since it is proportional to $\bar G_{AB}$, which would imply
that $\beta^2 = \gamma^2$ if the only contribution to $T$ came from the
cosmological constant on the brane. Thus the deviation of $T$ from the
$\bar G_{AB}$ form is due entirely to the contribution from the matter
Lagrangian. In a subsequent section we will show that the contribution to
the stress tensor from the zero point energies of fields living on the
brane lead to $\beta\ne\gamma$. In general $\beta$ and $\gamma$ will
depend on the geometrical factors $a,b$ and $c_2$ due to the quantum
contribution to the stress tensor from matter localized on the brane.

Now the Einstein equations for the upper 5 by 5 block of the Ricci tensor  
get modified in the presence of the branes to

\begin{equation}
2 M_*^4 (R_{AB} - g_{AB}R) = -\frac{1}{2} g_{AB} \Lambda_B 
+ 8 \pi \delta (r-a) T^a_{AB} + 8 \pi \delta (r-b) T^b_{AB}\ .
\end{equation}

Since we already have the solution in the bulk we can get the complete
solution by matching across the branes. The metric tensor is continuous
across the branes but its derivatives are not. The above equation fixes 
the jump discontinuity in the derivatives across the boundary.

In terms of components the conditions on the derivatives are

\begin{eqnarray}
-\frac{3}{2}\Delta\frac{f'}{f} - \frac{1}{2}\Delta\frac{s'}{s} &=& \beta^2
\\
2\Delta\frac{f'}{f} &=& -\gamma^2\ .
\end{eqnarray}

We wish to apply these conditions to our solution. To simplify matters we
first define

\begin{eqnarray}
\bar{\gamma}^2 &=& 4\beta^2-3\gamma^2 \\
F(c,r,\alpha) &=& \frac{2 \alpha c} {\left( e^{\frac{5}{2}r\alpha} - c
\right)}\ .
\end{eqnarray}

The conditions on the derivatives at the first boundary are
\begin{eqnarray}
\alpha_2 - \alpha_1 + F(c_2,a,\alpha_2) - F(-1,a,\alpha_1) &=& 
-\frac{1}{2}\gamma_1^2 \\
\alpha_2 - \alpha_1 -\frac{3}{2}\left[ F(c_2,a,\alpha_2) - F(-1,a,\alpha_1)\right]
+\frac{5}{2} \left[ F(-c_2,a,\alpha_2) - F(1,a,\alpha_1)\right] &=&
-\frac{1}{2}\bar{\gamma}_1^2 \ .
\end{eqnarray}

The conditions at the second boundary are 
\begin{eqnarray}
\label{eq:gam}
-\alpha_3 - \alpha_2 - F(c_2, b,\alpha_2) &=&-\frac{1}{2}\gamma_2^2 \\
\label{eq:gamb}-\alpha_3 - \alpha_2 +\frac{3}{2} F(c_2, b,\alpha_2) 
- \frac{5}{2} F(-c_2,b,\alpha_2) &=& -\frac{1}{2}\bar{\gamma_2}^2\ . 
\end{eqnarray}

Looking at the above equations we have three parameters $c_2$, $a$ and $b$
which have to satisfy four independent equations. Hence a fine tuning is
necessary.  This is the fine tuning necessary to set the effective
four dimensional cosmological
constant to zero. Once this fine tuning has been made it is
straightforward to find solutions to the above equations without
singularities where all
parameters are of order one in terms of the fundamental scale $M_*$.
However since we want to generate a hierarchy we want $\alpha_2 b =
O(40) \gg 1$.
From now on we will assume that $b$ is the only large parameter in the
problem and that it is therefore responsible for generating the hierarchy.
Subtracting eqn.~(\ref{eq:gam}) from eqn.~(\ref{eq:gamb}) we see that

\begin{equation}
\frac{5}{2} F(c_2, b,\alpha_2) - \frac{5}{2} F(-c_2,b,\alpha_2) =
-\frac{1}{2}\bar{\gamma_2}^2 +\frac{1}{2}\gamma_2^2\ .
\end{equation}

Note that from the definition of $\bar{\gamma}$ it is clear that if $\beta
= \gamma$, then $\bar{\gamma} = \gamma$.  But the difference between
$\beta$ and $\gamma$ arises from the vacuum energy of quantum fields
localized to the brane (the Casimir effect), which vanishes in the limit
of large proper radius for the compact dimension. As will be discussed in
a subsequent section the Casimir effect is finite and regulator
independent in the limit that the cutoff is taken to infinity.  Then by
dimensional considerations if the fields on the brane are massless the
right hand side must be of order $[2\pi s(b)]^{-\frac{5}{2}}$ since the proper size of the compact dimension
is the only scale in the problem. Then the equation above becomes

\begin{equation}
\frac{5}{2} F(c_2, b,\alpha_2) - \frac{5}{2} F(-c_2,b,\alpha_2) =
const [ e^{-\frac{5}{2}\alpha_2 b}(1 - c_2
e^{-\frac{5}{2}{\alpha_2}b})^{3}
(1 + c_2 e^{-\frac{5}{2}{\alpha_2}b})^{-5}]\ ,
\end{equation}
where the constant above is of order one in units of the fundamental
scale. While we see that both sides of this equation are the same order of
magnitude even for large $b$  we also see that $b$ only appears in the combination
$e^{-\frac{5}{2}\alpha_2 b}$. It is this combination which is determined in
terms of the various tensions. Hence although
the brane spacing is fixed,  
a large value for $\alpha_2 b$ is only possible through a
fine tuning  of parameters which is over and above the fine tuning necessary
to set the four dimensional cosmological constant to zero. In the next
section we shall show that by adding a scalar field in the bulk \`a la
Goldberger and Wise \cite{gw} the matching conditions at the branes can be 
made more sensitive to $b$ and the
hierarchy can be made natural\footnote{Note that the necessary 
fine tuning of the Planck
4-brane tension might also be  natural if the 4-brane is an
approximate BPS state of a nearly supersymmetric theory. Thus supersymmetry
of the bulk action  might provide an alternative solution to the
hierarchy problem. 
}.

Equating the various components of the metric across the boundary 
fixes the values of the coefficients $f_{02}$,  $s_{02}$, $f_3$,
and $s_3$. Once again there are four equations but there are now 
four unknowns so there is no further fine tuning required.

\begin{figure}[t]
\centerline{\epsfxsize=3 in \epsfbox{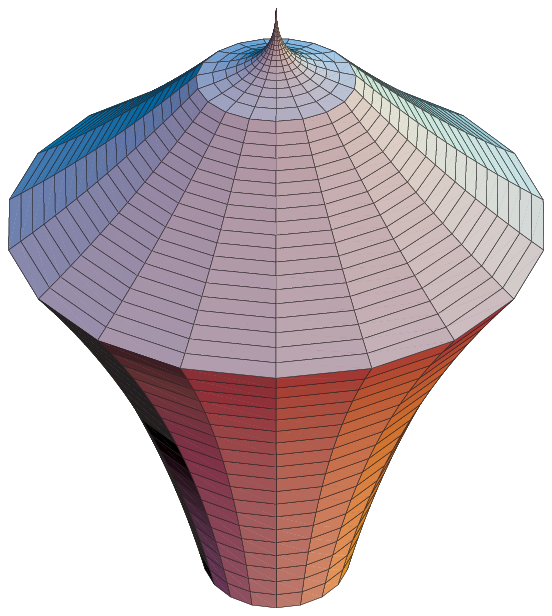}}
\vskip .3in
\noindent
Fig 1. Plot of warp factor of compact dimension (circumference of circles)  versus r (vertical axis), 
illustrating why we
call this the `space needle' metric. The apparent singularity at the
top, where $r=0,$ is  a coordinate singularity only.

\end{figure}

This completes the determination of the metric in the absence of any
bulk matter. The warp factor for the fifth dimension is plotted in
figure~1 for a choice of parameters for which there is a large hierarchy.
 The warp factor for the usual 3+1 dimensions is qualitatively similar,
except near $r=0$ where it goes to a constant.  We call this the ``space
needle metric'' for reasons which are obvious from the picture. 

\subsection{Model with Bulk Scalar Fields}

In this section we show that when the model of the previous section
is modified by the inclusion of bulk scalar fields, the hierarchy can
be made natural.  

The action for the scalar field
\begin{equation}
S_M =  \int d^6 x \sqrt{-G} \frac{1}{2}
(-\partial_M \psi \partial^M \psi - m^2 \psi^2) +
\sqrt{-\bar{G}}F(\psi) \delta(r-a) + \sqrt{-\bar{G}}H(\psi)
\delta(r-b)\ .
\end{equation}  

Here $m$ is the mass of the scalar field in the bulk. We will be 
interested in the limit $m^2 \ll \alpha^2$, since this is where we
naturally obtain a large hierarchy.
The scalar field sources $F(\psi)$ and $H(\psi)$ are in general arbitrary functions. 
For simplicity we will take   

\begin{eqnarray}
F(\psi) &=& \lambda_1 \psi \\
H(\psi) &=& \lambda_2 \psi\ . 
\end{eqnarray}

The coupled equations for the gravity-matter system are difficult to
solve. We shall address the problem by a successive approximation method.
We will first solve for the gravitational field in the absence of the
scalar field, assuming that the contribution to the stress tensor of this
field is small. We will then solve for the scalar field in this geometry,
and compute the stress tensor $T_{\psi}$ of the scalar field. We then can
substitute this solution back into the Einstein equations to determine the
correction to the geometry induced by the stress tensor of the scalar
field\footnote{We neglect quantum corrections to the stress tensor from
the scalar field, as we do not expect these to qualitatively alter our
conclusions.}. If we wish, this procedure, which is essentially an
expansion in $T_{\psi}/{M_*}^6$, can be carried out to higher orders but
the lowest order will be sufficient to to fix the brane spacing, which we
expect will receive only small corrections at higher orders.

For simplicity we will take as a starting metric

\begin{eqnarray}
f_0 &=& 1 \;\;\;\;\; r<a \\
f_0 &=& e^{-\alpha a} e^{\alpha r} \;\;\;\; a<r<b \\
f_0 &=&  e^{-\alpha a} e^{2\alpha b} e^{-\alpha r}\;\;\;\; r>b\ ,
\end{eqnarray}

\begin{eqnarray}
s_0 &=& r^2 \;\;\;\;\; r<a \\
s_0 &=& a^2 e^{-\alpha a} e^{\alpha r} \;\;\;\; a<r<b \\
s_0 &=& a^2 e^{-\alpha a} e^{2\alpha b} e^{-\alpha r}\;\;\;\; r>b\ .
\end{eqnarray}

This corresponds to a solution for the special case with no bulk
cosmological constant for $r<a$ but the same cosmological constant
everywhere outside, and $c_2 = 0$. This metric is flat in the neighborhood
of the origin and pure AdS outside $a$. This is a special case, with $\alpha_1$ and
$c_2$ set to zero, of the
more general class of metrics we have considered in section IIIA.  
Non-zero $c_2$  will be treated perturbatively in what
follows and we will also indicate how to include $\alpha_1$ perturbatively.
 Hence the only loss of generality arising from
this starting metric is  that we can only extend our 
conclusions to the more general class of metrics of the previous section 
when $c_2$ and $\alpha_1$ are sufficiently small that a perturbative
approach is valid.  

We first solve for the  scalar field in this background metric. 
The equation of motion   is

\begin{equation}
- \psi'' - \left(2 \frac{f'}{f} + \frac{1}{2} \frac{s'}{s}\right)
 \psi' + m^2 \psi
= \lambda_1 \delta(r-a) + \lambda_2 \delta(r-b)\ .
\end{equation}

The solutions in the three regions consistent with smoothness of $\psi$
at the origin and vanishing of $\psi$ at infinity are 

\begin{eqnarray}
\psi &=& A_1 \left(1 + \frac{1}{4} m^2 r^2 + \ldots\right) \;\;\; r<a \\
\psi &=& A_2 e^{\sigma_1 r} +  B_2 e^{\sigma_2 r} \;\;\; a<r<b \\
\psi &=& A_3 e^{\sigma_3 r} \;\;\; r>b\ ,
\end{eqnarray} 
where we have neglected higher order terms in $m^2r^2$ in equation (45),
and
\begin{eqnarray}
\sigma_{1} &=& -\frac{5}{4}\alpha - \sqrt{\left(\frac{5}{4}\alpha\right)^2 + m^2}\\
\sigma_{2} &=& -\frac{5}{4}\alpha + \sqrt{\left(\frac{5}{4}\alpha\right)^2 + m^2}\ .
\end{eqnarray}

For positive $m^2$ $\sigma_2$ is positive and $\sigma_1$
negative. Then 
\begin{equation}\sigma_3 = - \sigma_2\ .
\end{equation}  
The
coefficients $A_i$ and $B_i$ are to be determined by matching the
solutions for $\psi$ across the boundaries. We require continuity of
$\psi$ across the boundaries and the following jump conditions

\begin{equation}
\Delta \psi'(a) = -F'(\psi)
\end{equation}
at the first boundary and
\begin{equation}
\Delta \psi'(b) = -H'(\psi)
\end{equation}
at the second boundary. 

Since the expressions for the $A_i$'s and $B_i$'s are complicated we will
neglect effects of order $e^{-\alpha b}$ which are very small and further
assume $m^2 a^2 \ll 1$ so that such effects can also be neglected. These
approximations will not affect our conclusions.  Then in this limit the
expressions for the $A_i$'s and $B_i$'s are

\begin{eqnarray}
A_2 &=& \frac{\lambda_2 \sigma_2}{\left(\sigma_3 - \sigma_2\right) \sigma_1}
e^{(\sigma_2
- \sigma_1)a}e^{-\sigma_2b} -\frac{\lambda_1}{\sigma_1}  e^{-\sigma_1a}
\\
B_2 &=& -\frac{\lambda_2}{(\sigma_3 - \sigma_2)}e^{-\sigma_2b} \\
A_3 &=& -\frac{\lambda_2}{(\sigma_3 - \sigma_2)}e^{-\sigma_3b}\ .
\end{eqnarray}

    We now find the contribution to the stress tensor from the field
$\psi$, $T_{\psi}$. 

\begin{eqnarray}
16\pi{{T_{\psi}}^0}_0 &=& <-\frac{1}{2} ({\psi'}^2 + m^2) + F(\psi)
\delta(r-a)
+ H(\psi) \delta(r-b)> \\
16\pi{{T_{\psi}}^r}_r &=&  <\frac{1}{2} ({\psi'}^2 - m^2)>\ .
\end{eqnarray}

Next, we substitute the stress tensor back into the Einstein equations to
determine the corrections to the geometry. For convenience we define
$\overline{T} = 8\pi T_{\psi}/2{M_*}^4$, $\overline{\lambda} =
\lambda/2{M_*}^4$. The Einstein equations in the region $a<r<b$ take the
form

\begin{eqnarray}
-\frac{3}{2} f'' -\frac{1}{2} f \frac{s''}{s}
- \frac{3}{4}f' (\frac{s'}{s}) + \frac{1}{4}f
\left(\frac{s'}{s}\right)^2 &=& -\frac{5}{2}\alpha^2 f 
+ \overline{T_{00}} \\
2\frac{f''}{f} + \frac{1}{2}\left(\frac{f'}{f}\right)^2
&=& \frac{5}{2}\alpha^2  +  \overline{{T_{5}}^5} \\
\frac{3}{2}\left(\frac{f'}{f}\right)^2 +
\frac{f'}{f}\frac{s'}{s} &=& \frac{5}{2}\alpha^2  +
\overline{{T_{6}}^6}\ .
\end{eqnarray}

The equations inside $r<a$ can be obtained by setting $\alpha=0$ in the
equations above. The equations for $r>b$ are identical to those above.  We
are interested in the correction to the geometry to linear order in
$\overline{T}/{M_*}^2$.  To obtain this we expand

\begin{eqnarray}
f &=& f_0(1 + \epsilon) \\
s &=& s_0(1 + \kappa)\ ,
\end{eqnarray}
and linearize in $\epsilon$ and $\kappa$.  Then for $r<a$ the 
equations (58) and (60) above become 

\begin{eqnarray}
\epsilon' &=& \frac{1}{2}r \overline{{T^6}_6} \\
-\frac{3}{2} \epsilon'' - \frac{1}{2} \kappa'' - \frac{1}{r}\kappa'  
- \frac{3}{2r}\epsilon' &=& -\overline{{T^0}_0}\ .
\end{eqnarray}

These equations yield 

\begin{eqnarray}
\epsilon' &=& \frac{1}{2} r \overline{{T^6}_6} \\
\kappa' &=&  \frac{1}{2} r \overline{{T^6}_6} - 
\frac{5}{2} \frac{1}{r^2} \int_0^r dr r^2 \overline{{T^6}_6} \ ,
\end{eqnarray}
which can be integrated to get $\epsilon$ and $\kappa$. Here we have
used
the constraints of smoothness at the origin and
the linearized form of the Bianchi identity. 
The latter  is shown below.
\begin{equation}
(r\overline{{T^6}_6})' = \overline{{T^5}_5}\ . 
\end{equation}

For $a<r<b$ the equations (59) and (60) when linearized become

\begin{eqnarray}
2 \epsilon'' + 5 \alpha \epsilon' &=&  \overline{{T^5}_5} \\
4 \alpha \epsilon' + \alpha \kappa' &=&  \overline{{T^6}_6}\ .
\end{eqnarray}

These yield 

\begin{eqnarray}
\epsilon' &=& D e^{-\frac{5}{2}\alpha r} +
\frac{\overline{{T^6}_6}}{5\alpha} \\
\kappa' &=& -4 D e^{-\frac{5}{2}\alpha r} +
\frac{\overline{{T^6}_6}}{5\alpha}\ ,
\end{eqnarray}
which can be integrated to yield  $\epsilon$ and $\kappa$. 
Here $D$ is a constant of integration which must be determined 
from matching and once again the linearized form of the Bianchi 
identity for this region, shown below, has been used.

\begin{equation}
(\overline{T^6}_6)' + \frac{5 \alpha}{2} \overline{T^6}_6 
- 2 \alpha \overline{T^0}_0- \frac{1}{2} \alpha \overline{T^5}_5 
=0\ .
\end{equation}

In identical fashion we can get for the region $r>b$ 

\begin{equation}
\epsilon' = \kappa' = -\frac{\overline{{T^6}_6}}{5\alpha}\ .
\end{equation}

Here the requirement that the metric die away at infinity has been 
imposed. This completes the determination of bulk corrections to the
metric. The final step is to match across the branes thereby 
determining the brane positions and the undetermined constant $D$.

The conditions on the continuity of the metric across the brane are
straightforward to satisfy because of the additional constants of 
integration that will be obtained on integrating the expressions 
for $\epsilon'$ and $\kappa'$. Once $f$ has been normalized to one
on the visible brane and the requirement of smoothness has been 
met at $r=0$ all these additional constants will have been fixed.

We  now focus our attention on the jump conditions on the derivatives. 
At the inner brane we have

\begin{eqnarray}
-\frac{3}{2}\Delta \epsilon'(a) - \frac{1}{2}\Delta \kappa'(a) - 
\frac{1}{2}\left(\alpha - \frac{2}{a}\right) - \frac{3}{2}\alpha &=& {\beta_1}^2
- \frac{1}{2} \overline{\lambda_1} \psi(a) \\
2 \Delta \epsilon'(a) +2 \alpha &=& -{\gamma_1}^2 + \frac{1}{2}
\overline{\lambda_1} \psi(a)\ .
\end{eqnarray}

At the outer brane 

\begin{eqnarray}
-\frac{3}{2}\Delta \epsilon'(b) - \frac{1}{2}\Delta \kappa'(b) +
\frac{1}{2}(2\alpha) + \frac{3}{2}(2\alpha) &=& {\beta_2}^2
- \frac{1}{2} \overline{\lambda_2} \psi(b) \\
2 \Delta \epsilon'(b) -4 \alpha &=& -{\gamma_2}^2 + \frac{1}{2}
\overline{\lambda_2} \psi(b)\ .
\end{eqnarray}

These are four independent equations for the three unknowns $a,b$ and
$D$.
Just as in the previous section, the metric and brane locations are
completely
determined with one overall fine tuning needed to find a Poincar\'e 
invariant solution. 

First consider the situation without the scalar field, {\it i.e.} 
$\overline{T} =
0$.  This is the same  case  which was considered in the previous
section. In general, as we found, the metric beween the branes need
not be pure AdS even without a scalar field.  The tensions of the
branes must be finetuned  to set $D$ to zero. We are allowing
these brane tensions (which are related to
$\beta^2$ and $\gamma^2$ in the above equations) to deviate from those
fine tuned values by small amounts.
   Here small merely means that perturbation theory is valid.  Then it is
straightforward to verify that the equations (74) to (77) are simply
linearized versions of equations (29) to (32) with $\alpha_1 =0$ and the
parameter $D$ is proportional to $c_2$. Hence even in the absence of the
scalar field the metric between the branes is not of the AdS form, and our
conclusions once the scalar field is introduced will be valid for this
more general class of metrics.
 Setting $\overline{T}$ in the above formulas
to a constant non zero value for $r<a$ corresponds to allowing non zero
$\alpha_1$.  Although it is straightforward to accomodate non zero $\alpha_1$
perturbatively in this framework, for simplicity we shall keep
$\alpha_1=0$ in what follows below, and $\overline{T}$ will be related only to
$T_{\psi}$ and will not include any piece from a cosmological constant. 

We now return to the general case with a scalar field.  We are interested
in a solution with all parameters of order one in units of the fundamental
scale but we allow $m^2/\alpha^2$ to be somewhat less than one in order to
obtain a hierarchy.

Rather than give the exact solution 
we will give the more informative order of magnitude results. 
For dimensional
considerations any parameter of order one in units of the fundamental
scale will be denoted by the appropriate power of $\alpha$. 

The matching conditions at the inner brane, eqs. (74) and (75),
determine $a$ and $D$ as functions of $b$. Provided $m^2b/\alpha$ is
less than or of order one,  $a$ and $D$ are of  order one in units of
the fundamental scale. The  dependence of $D$ on
$b$  is  

\begin{equation} \label{eq:D}
D = O(\alpha) + O(m^2b)\ .
\end{equation}
This result   will turn out to be  crucial for the solution of
the hierarchy problem.

Both the equations at the outer brane are sensitive to
$b$ and $a$ only through exponentially small terms. 
This is because the value of $\psi$ and its derivatives is
exponentially insensitive to $b$ and $a$ in this region, which  manifests
itself in the forms of $\epsilon'$ and $\kappa'$ close to the outer brane. 
This is similar to the insensitivity to $b$ of equations (31) and
(32). One might therefore worry that just as in the previous case an
exponentially precise fine tuning will be necessary to get a
hierarchy. However the finetuning in the previous case was related to
the fact that near the Planck-brane the metric was very nearly pure
AdS, which is a homogenous space, and the compact dimension was 
exponentially large, so that the location of the Planck brane had very
little effect on the matching conditions at either brane. In the
present case with a light bulk scalar the value of the scalar at the
TeV brane  depends  more sensitively on $b$. To determine
$b$,
add equations (76) and (77) to obtain

\begin{equation}
\frac{5}{2} D e^{-\frac{5}{2}\alpha b} = (\beta_2)^2 - (\gamma_2)^2\ .
\end{equation}

This is the analogue of equation (33) in the previous section.
Recall that $(\beta_2)^2 - (\gamma_2)^2$ arises from the vacuum energy of
quantum fields localised on the brane and is of order $\alpha
e^{-\frac{5}{2}\alpha b}$. 
Then this equation, together with equation (\ref{eq:D})
which relates $D$ to $b$, determines $b$ as 

\begin{equation}
b = O\left(\frac{\alpha}{m^2}\right)\ .
\end{equation}

Clearly $m$ need not be much smaller than $\alpha$ to get a sizable
hierarchy. The difference of equations (76) and (77), which we have not
 yet used, 
becomes the condition that  the  effective four dimensional cosmological
constant is zero, which is the usual finetuning\footnote{Work
is in progress to see whether this finetuning can be made more natural
though supersymmetry of the bulk action when the Planck brane  is 
approximately BPS.}.

One limitation of the above approach is that the inner brane necessarily
has negative tension along the non compact directions. This is a
consequence of the choice of a starting metric with zero $\alpha_1$
and $c_2$ and
the fact that deviations from this metric can only be perturbative. The
 nonzero $\alpha_1$ case is not simple but there
is a limit in which it is tractable --- that in which $(\alpha_1 a)^2$ is 
perturbatively less than one, even though $\alpha_1$ is not small. 
In this limit, neglecting higher order
terms in  $(\alpha_1 r)^2$ we find 

\begin{eqnarray}
\frac{f'}{f} &=& \frac{5}{4} {\alpha_1}^2 r + \;.\;.\;. \\
\frac{s'}{s} &=& \frac{2}{r} - \frac{5}{6} {\alpha_1}^2 r +\; . \;. \;.
\end{eqnarray}

We can substitute for this in equation (44) and find that to the order
shown the solution (45) is unchanged. Now if $\alpha_1^2 a > \alpha_2$
the inner brane can have positive tension and it is straightforward to
verify that all the other conclusions above go through as before.

\subsection{Physical Implications}

To determine the effective four dimensional Planck scale we concentrate
on the higher dimensional Einstein action. When the two extra dimensions
are integrated out this will contain the usual four dimensional Einstein
action.

\begin{equation}
S_G = 2M_*^4 \int d^6x^M \sqrt{-G}R^6 \ .
\end{equation}

Expanding $g_{{\mu}{\nu}}=\eta_{{\mu}{\nu}} + h_{{\mu}{\nu}}$ and 
integrating out the two extra dimensions we see that

\begin{equation}
S_G = 2M_*^4 \int d^6x^M \sqrt{-g}R^4 f \sqrt{s} + \;\;\; \ldots
= M_4^2 \int d^4 x^{\mu} \sqrt{-g}R^4 + \;\;\; \ldots\ , 
\end{equation}
and consequently

\begin{equation}
2\pi M_*^4 \int dr f \sqrt{s} = M_4^2\ .
\end{equation}

Rather than do this integral exactly we will be satisfied with an 
estimate. The dominant contribution to the integral comes from the 
region close to the outer brane where the integrand is very large
and the forms of $f$ and $s$ are very nearly simple exponentials.
For the purposes of the estimate we set $\alpha_3 = \alpha_2 = 
a^{-1} = M_* = \alpha$. 

\begin{equation}
M_4^2 = O(\alpha^2 e^{\frac{3}{2}\alpha b})\ .
\end{equation}

This large exponent is responsible for the hierarchy between the Planck
scale and the weak scale. 

Next we analyze the spectrum of linearized tensor fluctuations to see if
it is consistent with the results of gravitational experiments. We neglect
the effect of the scalar field in what follows since we do not expect it
to qualitatively affect our results because its contribution to the energy
density is small. 

Expanding

\begin{equation}
G_{{\mu}{\nu}} = f\eta_{{\mu}{\nu}} + h_{{\mu}{\nu}}
\end{equation}
and substituting this into the Lagrangian we get the following equation 
for the fluctuation $h$ in the bulk.

\begin{equation}
-h'' - \frac{1}{2} \frac{s'}{s} h' - \left(\frac{f'}{f}\right)^2 h + 
\frac{5}{2} \alpha^2 h = \frac{1}{f}m^2 h\ ,
\end{equation}
where $m^2 = -\eta_{{\mu}{\nu}}p^{\mu}p^{\nu}$ and we are restricting our
attention to modes that have no momentum in the compact direction. These
are expected to be separated by a mass gap from the heavier modes with
nonzero momentum in the fifth direction. 

The boundary conditions that $h$ has to satisfy are $h'(0)=0$ at the
origin and 

\begin{equation} 
2\Delta\frac{h'}{h} = -\gamma^2\ .
\end{equation}

Clearly $h=f$ is a solution of this equation with $m=0$ by comparison with
eqns. (5) and (6). This is the massless graviton. There is a continuum of
other solutions for all positive $m^2$, as can be seen from the asymptotic
behavior of the equation. However to extract these solutions is not easy,
because of the complicated forms of $f$ and $s$. However since we are only
interested in order of magnitude estimates we can simplify the problem by
considering instead a simpler problem that retains the physics we are
interested in. Consider the metric 

\begin{eqnarray}
f &=& 1 \;\;\;\;\; r<a \\
f &=& e^{-\alpha a} e^{\alpha r} \;\;\;\; a<r<b \\
f &=&  e^{-\alpha a} e^{2\alpha b} e^{-\alpha r}\;\;\;\; r>b
\end{eqnarray}

\begin{eqnarray}
s &=& r^2 \;\;\;\;\; r<a \\
s &=& a^2 e^{-\alpha a} e^{\alpha r} \;\;\;\; a<r<b \\
s &=& a^2 e^{-\alpha a} e^{2\alpha b} e^{-\alpha r}\;\;\;\; r>b\ .
\end{eqnarray}

This corresponds to a solution for the case with no cosmological constant
for $r<a$ but the same cosmological constant everywhere outside. This
metric is flat in the neighborhood of the origin, AdS outside
$a$. It clearly has a form very similar to the metric we are interested in
for $r \gg a$, although unlike our case, the positions of the branes are
not fixed and one has negative tension. In what follows we will assume the
main features of the excitations we are interested in are common to both
metrics and proceed. 

With this approximation the equation in the region
 $ r<a$  where $\alpha$ is zero becomes

\begin{equation}
-h'' - \frac{1}{r} h' = m^2 h\ .
\end{equation}

The solution of this equation to leading order in $m^2 a^2$ is  

\begin{equation}
h = N \left[1 - \frac{1}{4} m^2 r^2\right]
\end{equation}
where $N$ is a normalization constant.  Since we are primarily interested
in the light modes this will suffice. In the region between the branes
the equation for the modes has the form

\begin{equation}
-h'' - \frac{1}{2} \alpha h' +
\frac{3}{2} \alpha^2 h = \frac{m^2}{f} h \ .
\end{equation}

This equation admits a solution in terms of Bessel functions. The 
solution is  

\begin{equation}
h = Ne^{-\frac{1}{4}\alpha r}( A_2 J_{\frac{5}{2}}[mq_2] + 
B_2 J_{-\frac{5}{2}}[mq_2])
\end{equation}
where $A_2$ and $B_2$ are constants and $q_2$ is defined by 

\begin{equation}
q_2 = \frac{2}{\alpha} \frac{ e^{-\frac{1}{2}\alpha r}}
{ e^{-\frac{1}{2}\alpha a}}\ .
\end{equation}

The closed form expressions for the relevant Bessel functions are 

\begin{eqnarray}
J_{\frac{5}{2}}(x) &=& \sqrt{\frac{2}{\pi x}}\left(\sin\;x\left[\frac{3}{x^2} - 1\right]
- 3\frac{\cos x}{x}\right) \\
J_{-\frac{5}{2}}(x) &=& \sqrt{\frac{2}{\pi x}}\left(\cos\;x\left[\frac{3}{x^2} - 1\right]
+3\frac{\sin x}{x}\right) \ .\\
\end{eqnarray}

Since we are interested in values of $m$ such that $mq_1 \ll 1$ we can 
conveniently approximate the solution between the branes by

\begin{equation} 
h =N \left[\bar{A}_2 \left(\frac{2m}{\alpha}\right)^{\frac{5}{2}} e^{-\frac{3}{2} \alpha
r} 
+ 3\bar{B}_2 \left(\frac{2m}{\alpha}\right)^{-\frac{5}{2}}e^{\alpha r}
+ \frac{1}{2} \bar{B}_2
\left(\frac{2m}{\alpha}\right)^{-\frac{1}{2}}e^{\alpha a}\right]\ .
\end{equation}

Coming to the region $r>b$ we can also obtain a solution in terms of 
Bessel functions.

\begin{equation}
h = Ne^{+\frac{1}{4}\alpha r}\left( A_3 J_{\frac{5}{2}}[mq_3] +
B_3 J_{-\frac{5}{2}}[mq_3]\right)
\end{equation}
where $q_3$ is defined by

\begin{equation}
q_3 = \frac{2}{\alpha} \frac{e^{\frac{1}{2}\alpha r}}
{ e^{-\frac{1}{2}\alpha a + \alpha b}}\ .
\end{equation}

The values of the $A$'s and $B$'s are to be determined by matching. The
boundary conditions to be satisfied are continuity of $h$ across the various
boundaries and the following jump conditions on the derivatives at $r=a$
and $r=b$ respectively
  
\begin{eqnarray}
\Delta\frac{h'}{h} &=& \alpha \\
\Delta\frac{h'}{h} &=& -2 \alpha\ .
\end{eqnarray}

The constant $N$ is to be determined by normalization.  Since this is not
quite in the form of a eigenvalue problem we make some simple
transformations which render it so.

Defining 

\begin{eqnarray}
g &=& const\;\;\; h \left(\frac{s}{f}\right)^{\frac{1}{4}} \\
q &=&  \int dr \frac{1}{\sqrt{f}}\ ,
\end{eqnarray}
we get an equation for $g$ as a function of $q$ which has the form of an
eigenvalue equation for $m^2$ with unit density function. For the 
continuum modes it is the outer region which is relevant for
normalization. But here $s$ and $f$ are 
proportional so we can conveniently choose the constant so that $g=h$ in 
this region. Also $q$ and $q_3$ as defined differ at most by an additive 
constant. Hence for the continuum modes we merely have to normalize 
the solution for h for $r>b$ with respect to $q_3$.

Matching at the inner brane we find that $\bar{A}_2$ is of
order $\sqrt{m/\alpha}$, and 
$\bar{B}_2$ is of order $(m/\alpha)^{\frac{5}{2}}$. Then matching at 
the outer brane we find that

\begin{eqnarray}
A_3 &=& O\left[\left(\frac{\alpha}{m}\right)^{\frac{1}{2}} 
e^{\alpha b}\right] \\
B_3 &=& O\left[\left(\frac{\alpha}{m}\right)^{-\frac{5}{2}} 
e^{-\frac{\alpha}{2}b}\right]\ .
\end{eqnarray}

In the far asymptotic region the $A_3$ mode is dominant. Normalizing to 
a box of size $L$ we find

\begin{equation}
N = O\left[\frac{1}{\sqrt L}\frac{m}{\alpha} e^{-\frac{3}{2}\alpha
b}\right] 
\ .
\end{equation}

The situation is slightly different for the zero mode. We write the
normalizable wave function as 

\begin{equation}
g_0 = N_0 f^{\frac{3}{4}} s^{\frac{1}{4}}\ .
\end{equation}

Now the integral relevant for normalization

\begin{equation}
\int dq g_0^2 = N_0^2 \int dr \frac{1}{\sqrt{f}} 
 f^{\frac{3}{2}} s^{\frac{1}{2}} = N_0^2 \int dr f s^{\frac{1}{2}} \ .
\end{equation}

This is the same integral that appears in determination of the four 
dimensional Planck scale. By exactly the same methods we obtain, on
normalizing to unity

\begin{equation}
N_0 = O(\alpha e^{-\frac{3}{4}\alpha b})\ .
\end{equation}  

We are now in a position to determine the corrections to gravity from the
Kaluza Klein excitations of the graviton. The change in the potential
energy between two masses $m_1$ and $m_2$ on our brane is given by

\begin{equation}
\Delta V = O\left[\frac{G m_1 m_2}{r} \int dm \left(\frac{N^2}{N_0^2 a}\right) L
e^{-mr}\right]
= O\left[G m_1 m_2 \frac{e^{-\frac{3}{2}\alpha b}}{\alpha^3 r^4}\right] =
O\left[\frac{G m_1 m_2}{r}\left(\frac{10^{-32}}{r^3
(TeV)^3}\right)\right]\ .
\end{equation}

 {} From this it is clear that deviations from Newtonian gravity are 
highly suppressed at long distances.

We now explain why we expect this model to have the same physical
implications as the model we started out with. Essentially for $r\gg a$ the
general solution of both models will have similar form. The
only difference will be in the magnitudes of the coefficients $\bar{A}_2$
and $\bar{B}_2$. Although these coefficients are determined by matching in
the interior their order of magnitude follows from simple dimensional
considerations. This then implies that $\bar{A}_3$ and $\bar{B}_3$ and
hence the normalization of the modes can be fixed by dimensional
considerations. Hence the two theories will give the same order of
magnitude estimate for the the corrections to Newtonian gravity. 

Because of the isometry of the compact dimension this theory contains
a massless ``gravi-photon''---a Kaluza-Klein $U(1)$ gauge
boson. However no light or massless fields will carry non-trivial U(1)
charge since they have no momentum in the compact dimension. We expect
that other
than the modes we have already discussed, the remaining spectrum of
gravitational excitations will be massive. 

A more comprehensive study of the phenomenological 
implications of this model is
left for future work.

\subsection{Stress Energy Tensor for a Field Localized to a Brane}

In this section we consider the form of the stress energy tensor for a
field localized to a brane having the metric $\bar{G}_{AB} =
diag(-1,1,1,1,1)$ but in which the fifth dimension is compact and has 
proper size $a$. We show
that in the ground state the stress tensor does indeed have the form given
in eqn.~(\ref{eq:stress}). For
simplicity we limit ourselves to the case of a free scalar field. 

The Lagrangian has the form

\begin{equation}
L = -\frac{1}{2}\partial_A \phi \partial^A \phi -  \frac{1}{2} m^2
\phi^2\ .  
\end{equation}

We are interested in the expectation value of the stress tensor in the
ground state.

\begin{equation}
<T_A^B> =  <-\partial_A \phi \partial^B \phi - \bar{G}_A^B L>\ .
\end{equation}

Because the ground state possesses translational symmetry

\begin{equation}
<T_A^B> = \frac{1}{V}\int d^4 x <T_A^B>
\end{equation}
where $V$ is the volume of the four space dimensions.  Performing a
Fourier expansion for the field $\phi$ and making use of the canonical
commutation relations this reduces to

\begin{eqnarray}
<T_0^0> &=& \sum_{\bf k} \sum_{k_5} \frac{1}{2V} 
\sqrt{{\bf k^2} + k_5 k^5 + m^2} \\
<T_m^n> &=& -\sum_{\bf k} \sum_{k_5}  \frac{1}{2V}
\frac{k_m k^n}{\sqrt{{\bf k^2} + k_5 k^5 + m^2}}\ .
\end{eqnarray} 

Since the usual three space dimensions are infinite the sums over momenta
in these three directions can be replaced by integrals. However since the
fifth dimension $\phi$ is compact the momenta in this direction remain
discrete, $k_5 = n/a$ where $n$ is an integer.

\begin{eqnarray}
<T_0^0> &=&  \sum_{k_5} \frac{1}{(2\pi)^4 a} \int d^3 {\bf k}
\sqrt{{\bf k^2} + k_5 k^5 + m^2} \\
<T_m^n> &=&  -\sum_{k_5}  \frac{1}{(2\pi)^4 a} \int d^3 {\bf k}
\frac{k_m k^n}{\sqrt{{\bf k^2} + k_5 k^5 + m^2}}\ .
\end{eqnarray}

These integrals are infinite and must be regulated in order to yield
sensible physical results. We will use a Pauli-Villars regulator, adding
massive fields with appropriate statistics until all the divergences have
been removed. (We could get similar results in a theory with
spontaneously broken supersymmetry.) We simplify to the special case where the
boson field is massless. Then all the divergences can be removed by adding
three fields with opposite statistics having masses $M$,$M$ and $2M$ and
two fields with the same statistics which both have mass $\sqrt{3}M$.
Here $M$ is assumed to be some kind of cut off for the theory. 

Performing the integrals and adding the contributions from the various 
fields we get the finite but regulator dependent results

\begin{eqnarray}
T_{\mu}^{\nu} &=& \delta_{\mu}^{\nu}  \sum_{k_5}
\frac{1}{4(2\pi)^3 a}
[k_5^4 ln(k_5^2) + 2(k_5^2 + 3 M^2)^2 ln(k_5^2 + 3 M^2) 
- 2(k_5^2 +  M^2)^2 ln(k_5^2 +  M^2) - (k_5^2 + 4 M^2)^2 ln(k_5^2 +
4M^2)]\\
T_{5}^{5} &=& -\sum_{k_5} \frac{1}{(2\pi)^3 a} k_5^2  
[k_5^2 ln(p_5^2) + 2(k_5^2 + 3 M^2) ln(k_5^2 + 3 M^2)
- 2(k_5^2 +  M^2) ln(k_5^2 +  M^2) - (k_5^2 + 4 M^2) ln(k_5^2 + 
4M^2)] \\
T_{\mu}^{5} &=& 0\ .
\end{eqnarray}

All other components of the stress energy tensor vanish, and it clearly
has the form of eqn (23). In a supersymmetric theory the scale $M$ will be
related to the scale of supersymmetry breaking.  We now estimate $T_m^n$
and $T_5^5$ in various limits. 

When $M \ll (1/a)$ we can approximate  $T_m^n$ and $T_5^5$ to leading
order in $(1/a)^2$ as $O(M^4/a)$ and $O(M^2/a^3)$ respectively. Clearly
in this limit $\beta^2$ and $\gamma^2$ are not equal.  

The relative difference between $T_5^5$ and $T^\nu_\mu$ must be finite
when the cutoff $M$ is taken to infinity and must vanish as the size of
the fifth dimension becomes infinite. We now estimate this difference as a
function of $a$ and $M$ when the dimensionless quantity $aM$ is large.

To do this we attempt to replace the sum we are interested in by a
sum of 
integrals. 
We begin by observing that the integral below can be broken up into a 
sum of integrals over equal subdomains.

\begin{equation}
\int dk^5 T(k^5) = \sum_{k_5} \int_{k_5}^{k_5+\frac{1}{a}} dp^5
T(p^5)\ .
\end{equation}

Here $T$ represents an arbitrary function of $k_5$. If the function $T$ 
is smooth it can be Taylor expanded 

\begin{equation}
\label{eq:taylor}
T(p^5) = T(k_5) +  T'(k^5) (p^5-k^5) + T''(k^5) \frac{(p^5-k^5)^2}{2} +
\ldots 
\end{equation}

Then performing the integrals over the subdomains we get

\begin{equation}
\int dk^5 T(k^5) = \sum_{k_5} \left[ \frac{1}{a}  T(k_5) + \frac{1}{2a^2}  
T'(k_5) + \frac{1}{6a^3} T''(k_5) +  \ldots \right]\ . 
\end{equation}

Now if $\sum T$ is $T_m^n$ or $T_5^5$ the first term on the right has the 
form we are interested in. Also notice that the other terms on the right 
then involve fewer powers of $k_5$ in the numerator and hence for 
the seventh term and beyond the sums for individual fields are finite
and straightforward to estimate.

$\sum T$ can also be chosen to be derivatives to arbitrary order of
$T_m^n$ or $T_5^5$. Since these quantities ocur in the expansions for
$T_m^n$ or $T_5^5$ they can then be substituted back to obtain systematic
expansions for $T_m^n$ and $T_5^5$ in terms of integrals. 

A complication that arises for the case of a massless field is that 
the fourth derivative and higher of $T_m^n$ and $T_5^5$ are not well
defined at $k_5 = 0$. We account for this by separating this point from 
the sum and approximating the rest of the sum by integrals from $(1/a)$
to infinity and $-(1/a)$ to negative infinity.

To obtain a reasonable estimate we must expand in eq.~(\ref{eq:taylor}) to
at least seventh order in order to account for all possible divergences as
powers or logarithms of $M$.  The calculation is straightforward but
lengthy and the details will not be presented here. The result is that the
difference between $T_m^n$ and $T_5^5$ is finite and of order $(1/a)^5$ in
the limit that the cutoff $M$ is large. 

This $(1/a)^5$ result for the difference could have been anticipated. 
Since the only
counterterm allowed by general covariance is the cosmological constant
which contributes equally to both $T_m^n$ and $T_5^5$, the difference
between these two must be finite and regulator independent in the limit
that the cutoff is taken to infinity. For a massless field, $a$
is the only available dimensionful parameter.  This result is just a
higher dimensional form of the Casimir force.

\section{Conclusions}

We have constructed a set of solutions to Einstein's equations in 
six or more dimensions, and exhibited a six dimensional set up, ``the
space needle'', with  
two concentric  positive tension 4-branes, which each have 
one compact dimension.  Gravity is mostly
localized
to the outer brane
while we assume the standard model lives on the inner brane,
explaining the apparent weakness of gravity in our world. There are no
massless moduli
associated with either the size of the compact dimension 
or the brane locations. This provides an explicit demonstration that
the gauge hierarchy problem can be solved in six dimensions, 
without supersymmetry,
and with negligible corrections to  gravity at  distances longer than
an inverse TeV.
Gravitational effects do become strong at energies of order a TeV. We
leave discussion of gravitational collider phenomenology of new
noncompact dimensions for future work.

We do not address the important issue of how to obtain chiral fermions
on the standard model brane. Ordinary dimensional reduction by
compactifying the fifth dimension on a circle always results in a
non-chiral theory.  One simple alternate possibility is to 
have the standard model live on a 3-brane at the center of space. 
Then it is only
necessary to have one 4-brane---the Planck brane. The metric is
simply the $a\rightarrow 0$ limit of the space needle metric. Alternatively,
it may be possible
to generalize our mechanism to additional dimensions with
some  compactification which does
allow a chiral effective theory on the TeV brane. 

We also do not address the cosmological constant problem. 
The effective four dimensional cosmological constant depends on a
complicated function of the  bulk and brane parameters, and may
be finetuned to zero or to a small  acceptable value.

\goodbreak
{\bf Acknowledgements}

This work was partially supported by the DOE under contract
DE-FGO3-96-ER40956. A.E.N. would like to acknowledge useful
conversations with D.B. Kaplan. Z.C. would like to acknowledge useful
conversations with S.P. Kumar and M. Luty.


\begin{thebibliography}{99}

\bibitem{add}
N.~Arkani-Hamed, S.~Dimopoulos and G.~Dvali,
Phys.\ Lett.\  {\bf B429}, 263 (1998)
[hep-ph/9803315],I.~Antoniadis, N.~Arkani-Hamed, S.~Dimopoulos and G.~Dvali,
Phys.\ Lett.\  {\bf B436}, 257 (1998)
[hep-ph/9804398].
\bibitem{rs}
L.~Randall and R.~Sundrum,
Phys.\ Rev.\ Lett.\  {\bf 83} (1999) 3370,
[hep-ph/9905221],
hep-th/9906064.
\bibitem{ck}
A.~G.~Cohen and D.~B.~Kaplan,
hep-th/9910132;N.~Arkani-Hamed, L.~Hall, D.~Smith and N.~Weiner,
hep-ph/9912453.
\bibitem{early}
 O.~Klein,
Z.\ Phys.\  {\bf 37}, 895 (1926), Nature {\bf 118}, 516 (1926);
K. Akama,  in Proceedings of the International Symposium on Gauge Theory
and Gravitation, (Springer-Verlag, 1983), 
edited by K. Kikkawa, N. Nakanishi and H. Nariai, 267-271, hep-th/0001113,  
Prog. Theor. Phys. {\bf 78} 184 (1987) ,{\it op. cit.} {\bf 79} 1299 (1988);
M.~Visser,
Phys.\ Lett.\  {\bf B159}, 22 (1985); 
M.~Gell-Mann and B.~Zwiebach,
Nucl.\ Phys.\  {\bf B260}, 569 (1985);
V.~A.~Rubakov and M.~E.~Shaposhnikov,
Phys.\ Lett.\  {\bf B125}, 136, 139 (1983);
I.~Antoniadis,
Phys.\ Lett.\  {\bf B246}, 377 (1990);
J..~D.~Lykken,
Phys.\ Rev.\  {\bf D54}, 3693 (1996)
[hep-th/9603133];
R.~Sundrum,
JHEP {\bf 9907}, 001 (1999)
[hep-ph/9708329].
\bibitem{rl}
J.~Lykken and L.~Randall,
hep-th/9908076. See also
N.~Arkani-Hamed, S.~Dimopoulos, G.~Dvali and N.~Kaloper,
hep-th/9907209.
\bibitem{cosmo}
C.~Csaki, M.~Graesser and J.~Terning,
Phys.\ Lett.\  {\bf B456}, 16 (1999)
[hep-ph/9903319];
P.~Binetruy, C.~Deffayet and D.~Langlois,
hep-th/9905012;
C.~Csaki, M.~Graesser, C.~Kolda and J.~Terning,
Phys.\ Lett.\  {\bf B462}, 34 (1999)
[hep-ph/9906513]; 
J.~M.~Cline, C.~Grojean and G.~Servant,
Phys.\ Rev.\ Lett.\  {\bf 83}, 4245 (1999)
[hep-ph/9906523];
D.~J.~Chung and K.~Freese,
hep-ph/9906542;
T.~Shiromizu, K.~Maeda and M.~Sasaki,
gr-qc/9910076;
C.~Csaki, M.~Graesser, L.~Randall and J.~Terning,
hep-ph/9911406;
W.~D.~Goldberger and M.~B.~Wise,
hep-ph/9911457;
P.~Binetruy, C.~Deffayet, U.~Ellwanger and D.~Langlois,
hep-th/9910219;
P.~Kanti, I.~I.~Kogan, K.~A.~Olive and M.~Pospelov,
hep-ph/9912266.
\bibitem{gw}
W.~D.~Goldberger and M.~B.~Wise,
Phys.\ Rev.\ Lett.\  {\bf 83}, 4922 (1999)
[hep-ph/9907447].
\bibitem{stabilize}
N.~Arkani-Hamed, S.~Dimopoulos and J.~March-Russell,
hep-th/9809124;
R.~Sundrum,
Phys.\ Rev.\  {\bf D59}, 085010 (1999)
[hep-ph/9807348];
M.~A.~Luty and R.~Sundrum,
hep-th/9910202.
\bibitem{cp} A. Chodos and E. Poppitz, Phys. Lett. {\bf B471}, 119
(1999)
[hep-th/9909199].
\end{thebibliography}
\end{document}